\documentclass[aps,prb,twocolumn]{revtex4}
\usepackage{graphicx}
\begin{document}

\title{A ``partitioned leaping" approach for multiscale modeling of chemical reaction dynamics}

\author{Leonard A. Harris}
\email{lh64@cornell.edu}%
\affiliation{School of Chemical and Biomolecular Engineering,
Cornell University, Ithaca, NY 14853, USA}%

\author{Paulette Clancy}%
\email{pqc1@cornell.edu}%
\affiliation{School of Chemical and Biomolecular Engineering,
Cornell University, Ithaca, NY 14853, USA} %

\date{\today}

\begin{abstract}
We present a novel multiscale simulation approach for modeling
stochasticity in chemical reaction networks.  The approach
seamlessly integrates exact-stochastic and ``leaping"
methodologies into a single {\em partitioned leaping\/}
algorithmic framework. The technique correctly accounts for
stochastic noise at significantly reduced computational cost,
requires the definition of only three model-independent parameters
and is particularly well-suited for simulating systems containing
widely disparate species populations. We present the theoretical
foundations of partitioned leaping, discuss various options for
its practical implementation and demonstrate the utility of the
method via illustrative examples.
\end{abstract}

\maketitle

\section{Introduction}

Stochastic simulations of chemical reaction networks have become
increasingly popular recently, largely due to the observation that
stochastic noise plays a critical role in biological
function.~\cite{McAdams97, Arkin98, McAdams99, Elowitz02, Fedo02,
Rao02, Kaern05, Raser05}  The issue is relevant in other
scientific fields as well, such as microelectronics processing,
where statistical variations in dopant profiles can profoundly
affect the performance of nanoscale semiconductor
devices.~\cite{Plumm95, Roy05, ITRS01}  Gillespie's stochastic
simulation algorithm (SSA),~\cite{Gillesp76} in particular, has
found widespread use in computational biology. The method is a
kinetic Monte Carlo approach that produces time-evolution
trajectories correctly accounting for the inherent stochasticity
associated with molecular interactions. Detailed accuracy is
achieved by explicitly simulating every reaction occurrence within
a system. The method is computationally expensive as a result,
however, and practical application is limited to only very small
systems.

Motivated by this fact, considerable effort has been undertaken
recently to develop accelerated simulation methods capable of
correctly accounting for stochastic noise but at significantly
reduced computational cost. Broadly speaking, these endeavors can
be divided into three categories: (i) algorithmic advances to
increase the efficiency of exact-stochastic
methods,~\cite{Gibson00, Cao04:efficientSSA, Resat01} (ii)
``leaping" techniques in which efficiency is enhanced by ignoring
the exact moments at which reaction events occur,~\cite{Gillesp01,
Gillesp03, Rath03, Cao04:stability, Cao05:negPop, Rath05,
Cao06:newStep, Tian04, Chatt05} and (iii) ``partitioned" methods
in which sets of reactions are divided into various
classifications, such as ``fast" and ``slow," and treated either
by applying appropriate numerical techniques to each subset
(``methodology coupling" or ``hybrid")~\cite{Hasel02, Burr04,
Puch04, Vasu04, Kiehl04, Salis05:hybrid} or by reducing the model
to incorporate the {\em effects\/} of the fast reactions into the
dynamics of the slow (``model reduction").~\cite{Rao03,
Cao05:slowSSA, Cao05:MSSA, Gout05, Samant05, Salis05:QSSA}

In this article, we present a new simulation method for modeling
chemical reaction dynamics that integrates exact-stochastic,
leaping and methodology coupling methods into a single multiscale
algorithmic framework. The fundamental theory underlying
Gillespie's $\tau$\/-leap approach~\cite{Gillesp00, Gillesp01} is
used to formulate a theoretically justifiable partitioning scheme.
The partitioning is based on the number of reaction firings
expected within a calculated time interval and partitioning and
time step determination are thus inextricably linked.  Based on
the value of the time step each reaction is then ``classified" at
one of various levels of description. Species populations are
updated for coarsely classified reactions using the leaping
formulas introduced by Gillespie,~\cite{Gillesp01} while rare
events are treated via the methods of Gibson and Bruck's
exact-stochastic Next Reaction method.~\cite{Gibson00} Overall,
the technique efficiently simulates systems containing widely
disparate species populations using rigorously derived
classification criteria and requiring minimal user intervention.

We begin in Sec.~\ref{sec:bkgrd} by reviewing the theoretical
foundations of exact-stochastic simulation and $\tau$\/-leaping.
This provides a basis for discussing partitioned leaping in
Sec.~\ref{sec:PLA} as well as various options for its practical
implementation. In Sec.~\ref{sec:examples} we present two
illustrative examples demonstrating the utility of the method: one
inspired by biology and a clustering example relevant to materials
and atmospheric sciences. Finally, in Sec.~\ref{sec:discuss} we
summarize the attributes of the method, discuss possible
modifications, and draw conclusions regarding its place among the
many alternative techniques that have been proposed.

\section{Background} \label{sec:bkgrd}

As is customary, we consider a well-mixed system of $N$\/
molecular species $\{S_1, \ldots, S_N\}$ interacting via $M$\/
reaction channels $\{R_1, \ldots, R_M\}$ in a volume $\Omega$ at
constant temperature.  The state of the system is described by the
vector $\mathbf{X}$($t$\/), where $X_i(t)$ represents the
population of species $S_i$\/ at time $t$\/.  Each reaction
channel $R_\mu$\/ has associated with it a propensity function
$a_\mu$\/ and a state-change (or stoichiometric) vector
$\mathbf{z}_\mu = (z_{{\mu}1}, \ldots,
z_{{\mu}N})$.~\cite{note:subscr} The propensity $a_\mu$\/ is
defined such that $a_{\mu}dt$\/ gives the probability that
reaction $R_\mu$\/ will fire once during the infinitesimal time
interval $dt$\/.  In other words, $a_\mu$\/ is the stochastic
analog to the deterministic {\em reaction rate\/}.  As such,
$a_\mu$\/ can be written as the product of a stochastic rate
constant $c_\mu$\/ (which is related to the deterministic rate
constant via a simple scaling by the system
volume)~\cite{Gillesp76} and a combinatorial factor $h_\mu$\/,
which represents the number of distinct ways in which a
realization of $R_\mu$\/ can occur.  In general, $h_\mu$\/ is a
simple function of the reactant species populations for
$R_\mu$.~\cite{Gillesp76}

\subsection{Exact stochastic simulation} \label{subsec:ES}

Given the definitions above, Gillespie has developed a simulation
methodology for modeling chemical reaction dynamics that
``exactly" accounts for the stochastic nature of the
process.~\cite{Gillesp76} The stochastic simulation algorithm, or
SSA, is exact in the sense that it produces {\em possible\/}
time-evolution trajectories that are consistent with the
underlying {\em chemical Master Equation\/} governing the physical
process.~\cite{Gillesp92}

The SSA operates by generating, at each simulation step, random
samples of reaction times ($\tau$\/) and types ($\mu$\/) and
advancing the system forward in time accordingly.  Two
alternative, yet equivalent, strategies exist for doing so.  The
first, dubbed the {\em Direct method\/} (DM), generates $\tau$\/
according to~\cite{Gillesp76}
\begin{equation}
    \tau = -\ln(r_1) / a_0,
    \label{eq:Direct-tau}
\end{equation}
while $\mu$\/ is the integer satisfying the relationship
\begin{equation}
    \sum_{\nu=1}^{\mu-1} a_\nu < a_0 r_2 \leq
    \sum_{\nu=1}^\mu a_\nu,
    \label{eq:Direct-mu}
\end{equation}
where $a_0 \equiv \sum_\nu a_\nu$\/ and $r_1$ and $r_2$ are
unit-uniform random numbers between 0 and 1.

The second approach considers each reaction in the system on an
{\em individual\/} basis and asks the question, ``When will
reaction $R_\mu$ next fire {\em assuming\/} that no other
reactions fire first?"  Values of such ``tentative" next-reaction
times, $\tau_\mu^\mathrm{ES}$,~\cite{note:label} are then
generated for each reaction. $\tau$\/ is set equal to the smallest
of these and $\mu$\/ to the corresponding reaction since this is
the only one for which the assumption that no other reactions fire
first actually holds. This approach was originally developed by
Gillespie~\cite{Gillesp76} and dubbed the {\em First Reaction
method\/} (FRM).  Tentative next-reaction times are calculated via
\begin{equation}
        \tau_\mu^\mathrm{ES} = -\ln(r_\mu) / a_\mu,
        \label{eq:tau-first}
\end{equation}
where $r_\mu$\/ is a unit-uniform random number.

This approach has recently been modified by Gibson and
Bruck.~\cite{Gibson00} The Next Reaction method (NRM) is more
computationally efficient in that a rigorous transformation
formula is employed that significantly reduces the number of
random number generations required during the course of a
simulation.  Once a reaction has fired, a new value of
$\tau_\mu^\mathrm{ES}$ is generated {\em for that reaction only\/}
using Eq.~(\ref{eq:tau-first}).  For all other reactions
\begin{equation}
    \tau_\mu^\mathrm{ES} = (a_\mu^\prime / a_\mu)
    (\tau_\mu^{\prime\,\mathrm{ES}} - \tau^\prime),
    \label{eq:tau-next}
\end{equation}
is used, where the unprimed and primed quantities signify new and
old values, respectively.~\cite{note:NRM-differ}

It should be noted that a recent timing
analysis~\cite{Cao04:efficientSSA} has shown that, while the NRM
is certainly more efficient than the FRM, an optimized version of
the DM actually performs best in most situations.  For our
purposes, however, this fact is not important. The {\em ideas\/}
underlying the FRM are what we will use in the development of our
new simulation approach, and the increased efficiency offered by
the NRM will be utilized in its implementation.

\subsection{$\tau$\/-leaping} \label{subsec:tauLeap}

Despite the improved efficiency offered by the NRM~\cite{Gibson00}
and the optimized version of the DM,~\cite{Cao04:efficientSSA} the
SSA remains limited as to the system size amenable to treatment
due to its ``one reaction at a time" nature.  As a result,
Gillespie has recently attempted to move beyond the ``exact"
approach by introducing approximations regarding the number of
times a reaction can be expected to fire within a given time
interval. His approach, known as $\tau$\/-leaping, begins by
defining the quantity $K_\mu(\tau)$ as the number of times
reaction channel $R_\mu$\/ fires during the time interval
$[t,t+\tau)$.~\cite{Gillesp00, Gillesp01} In general,
$K_\mu(\tau)$ is a complex random variable dependant upon the
propensity $a_\mu$\/ and the manner in which it changes during
$[t,t+\tau)$. Obtaining a rigorous expression for the probability
function governing $K_\mu(\tau)$ is thus tantamount to solving the
usually intractable chemical Master Equation.

Gillespie recognized, however, that if a time period exists over
which the propensity $a_\mu$\/ remains {\em essentially\/}
constant then one can approximate $K_\mu(\tau)$ as a {\em Poisson
random variable\/},~\cite{Gillesp00, Gillesp01}
\begin{equation}
    K_\mu(\tau) \approx \mathcal{P}_\mu(a_\mu,\tau).
    \label{eq:K-poisson}
\end{equation}
There always exists a value of $\tau$\/ over which this assumption
holds; in the extreme case it is the interval between successive
reactions. In many cases, however, the interval is likely to span
numerous reaction events, especially when the reactant populations
are ``large."~\cite{Gillesp01}  Gillespie then noted that if the
mean value of $\mathcal{P}_\mu(a_\mu,\tau)$, i.e.,
$a_\mu{\tau}$\/, is ``large," then one can approximate the Poisson
random variable as a {\em normal\/} random
variable,~\cite{Gillesp00, Gillesp01}
\begin{eqnarray}
    K_\mu(\tau) & \approx & \mathcal{N}_\mu(a_\mu\tau,a_\mu\tau)
    \label{eq:K-norm} \\
    & = & a_\mu\tau + \sqrt{a_\mu\tau} \times
    \mathcal{N}(0,1), \nonumber
\end{eqnarray}
where the second equality follows from the linear combination
theorem for normal random variables.~\cite{Gillesp00, Gillesp01}
Equation~(\ref{eq:K-norm}) is essentially a chemical Langevin
equation and amounts to a ``continuous-stochastic" representation
of the reaction dynamics.  Finally, Gillespie showed that if the
ratio of the ``deterministic" term in (\ref{eq:K-norm}),
$a_\mu{\tau}$\/, to the ``noise" term, $\sqrt{a_\mu{\tau}}$\/, is
``large," then the noise term can be neglected,
leaving~\cite{Gillesp00, Gillesp01}
\begin{equation}
    K_\mu(\tau) \approx a_\mu\tau,
    \label{eq:K-determ}
\end{equation}
or a ``continuous-deterministic" representation.

The expressions in Eqs.~(\ref{eq:K-poisson})--(\ref{eq:K-determ})
represent a theoretical ``bridge" connecting the
discrete-stochastic representation of reaction dynamics and the
more familiar continuous-deterministic description. With reactions
classified at one of these levels of description, these formulae
are used in $\tau$\/-leaping to determine the number of times each
reaction fires within a given time interval.  One is thus able to
``leap" forward in the temporal evolution of a system multiple
reaction firings at a time.

Implementing these ideas algorithmically requires a practical
method for determining the time interval over which the Poisson
approximation (\ref{eq:K-poisson}) can be expected to hold.
$\tau$\/-selection has been the subject of extensive
research~\cite{Gillesp01, Gillesp03, Cao06:newStep} and has
undergone numerous refinements recently.  The basic idea is to
impose a constraint on the magnitude of the change of an
individual reaction propensity,
\begin{equation}
    \left|\,a_\mu(t+\tau_\mu^\mathrm{Leap})-a_\mu(t)\,\right| / \xi =
    \epsilon , \,\,\,(0 < \epsilon \ll 1) ,
    \label{eq:constraint}
\end{equation}
where $\xi$\/ is an appropriate scaling factor. In applying this
constraint, one seeks to identify the time interval
$\tau_\mu^\mathrm{Leap}$\/ over which the propensity $a_\mu$\/ for
reaction $R_\mu$\/ will remain essentially constant within a
factor of $\epsilon$\/ {\em assuming\/} that the propensities for
all other reactions also remain essentially constant. One then, in
essence, calculates a value of $\tau_\mu^\mathrm{Leap}$ for each
reaction in the system and sets $\tau$\/ equal to the smallest of
these.

We will discuss $\tau$\/-selection further in
Sec.~\ref{subsec:PLA-timestep}.  For now, however, note that there
is an interesting analogy between the procedure used in
$\tau$\/-leaping and that in the FRM (and NRM by extension). In
both cases, a time interval is calculated for a specific reaction
channel with assumptions made regarding all other reaction
channels. The time step is then set equal to the smallest of the
set as it is the only one for which the assumptions actually hold.
This suggests that the two methods might be seamlessly merged in
some way, and forms the basis of our new simulation approach.

With the time step $\tau$\/ calculated, one then proceeds to
determine the number of times each reaction fires in $\tau$\/
using Eqs.~(\ref{eq:K-poisson})--(\ref{eq:K-determ}). Strictly
speaking, the $\tau$\/-leap method uses only
Eq.~(\ref{eq:K-poisson}).  If the propensities of {\em all\/}
reactions are such that $\{a_\nu\tau\} \gg 1$, however, then
Eq.~(\ref{eq:K-norm}) is employed. In Ref.~\onlinecite{Gillesp01}
this is termed the {\em Langevin method\/}.  Furthermore, if all
$\{\sqrt{a_\nu\tau}\} \gg 1$ then Eq.~(\ref{eq:K-determ}) is
employed, which is equivalent to the explicit Euler method for
solving ordinary differential equations.~\cite{Gillesp01} Finally,
a {\it proviso\/} is added~\cite{Gillesp01} whereby the SSA is
used if $\tau \lesssim 1/a_0$, since $1/a_0$ is the expected time
to the next reaction firing in the system.~\cite{Gillesp76}

Additional modifications to $\tau$\/-leaping have been introduced
recently.  Primarily, strategies have been developed to prevent
the possible occurrence of negative species populations. This
possibility arises from the fact that Poisson and normal random
variables are positively unbounded and, while unlikely, could
produce physically unrealizable numbers of reaction firings that
result in the consumption of more reactant entities than are
present in the system.  Tian and Burrage~\cite{Tian04} and
Chatterjee~{\it et al.}~\cite{Chatt05} avoid this problem by using
binomial random numbers rather than Poisson.  Cao~{\it et
al.}~\cite{Cao05:negPop} identify ``critical" reactions deemed in
danger of exhausting their available reactant populations and
treat them using a DM SSA approach.  The interested reader is
referred to Ref.~\onlinecite{Cao06:newStep} for further details
regarding the current state of $\tau$\/-leaping methodologies.

\section{Partitioned leaping} \label{sec:PLA}

\subsection{The Framework} \label{subsec:PLA-algo}

The primary change that we make to $\tau$\/-leaping involves
classifying reactions {\em individually\/} once the time step
$\tau$\/ has been determined.  The expressions
(\ref{eq:K-poisson})--(\ref{eq:K-determ}) are derived for
individual reactions and there is nothing precluding their use in
this manner.  Thus, sets of reactions can essentially be
partitioned into ``fast," ``medium," and ``slow" classifications
based on the quantities $\{a_\nu\tau\}$. We can apply Gillespie's
{\it proviso\/}~\cite{Gillesp01} to each individual reaction as
well, essentially classifying a reaction as ``very slow" if $\tau
\lesssim 1/a_\mu$.  A tentative next-reaction time for such a
reaction can then be generated from Eqs.~(\ref{eq:tau-first}) or
(\ref{eq:tau-next}) and $R_\mu$\/ deemed to fire if
$\tau_\mu^\mathrm{ES} \leq \tau$.

This procedure amounts to a theoretically justifiable partitioning
scheme in which reactions are classified into four different
categories based on their propensity values, the calculated time
step $\tau$\/, and the criteria identified by
Gillespie~\cite{Gillesp00, Gillesp01} for transitioning between
the descriptions (\ref{eq:K-poisson})--(\ref{eq:K-determ}). The
classifications are made as follows:
\begin{itemize}
    \item{If $a_\mu\tau \lesssim 1 $\, $\rightarrow$ {\it
          Exact Stochastic\/} (very slow)}
    \item{If $a_\mu\tau > 1$ but $\not\gg 1$\,
          $\rightarrow$ {\it Poisson\/} (slow)}
    \item{If $a_\mu\tau \gg 1$ but
          $\sqrt{a_\mu\tau} \not\gg 1$\, $\rightarrow$ {\it
          Langevin\/} (medium)}
    \item{If $\sqrt{a_\mu\tau} \gg 1$\, $\rightarrow$ {\it
          Deterministic\/} (fast)}
\end{itemize} \label{list:classify}

These classifications constitute the basis of the partitioned
leaping approach.  At each simulation step, a time step $\tau$\/
is calculated and each reaction classified in the manner outlined
above. The numbers of reaction firings are then determined, based
on these classifications, using
Eqs.~(\ref{eq:tau-first})--(\ref{eq:K-determ}) and the system
evolved accordingly.

Various technical issues must be considered, however, in order to
correctly implement this approach. The first involves the
inclusion of the ``Exact Stochastic" (ES) classification and the
random nature of $\tau_\mu^\mathrm{ES}$. Specifically, if
$\tau_\mu^\mathrm{ES} < \tau$\/, and the clock is subsequently
advanced by $\tau$\/, then the possibility of $R_\mu$\/ firing
again within the interval $(\tau - \tau_\mu^\mathrm{ES})$ is
precluded. To overcome this complication we employ an iterative
procedure for determining $\tau$\/ and classifying reactions. Once
the reaction classifications are made $\tau_\mu^\mathrm{ES}$
values are calculated for all ES reactions. Some of these may be
smaller than $\tau$\/, and $\tau$\/ is thus adjusted to the
smallest of these. Decreasing $\tau$\/ will result in decreased
values of $\{a_\nu\tau\}$ and each reaction must thus be
reclassified. Some reclassified reactions may become ES which
previously were not and $\tau_\mu^\mathrm{ES}$ values must thus be
calculated for all of these reactions. Again, these may be smaller
than $\tau$\/ and the procedure is thus repeated until no further
adjustments are necessary.~\cite{note:ES-iter}

Situations may also arise in which all values of
$\tau_\mu^\mathrm{ES}$ are larger than $\tau$\/. In this case, we
leave $\tau$\/ unchanged {\em unless\/} all reactions in the
system are classified as ES.  This is because when all reactions
are ES it is always safe to allow $\tau$\/ to increase to the time
at which the next reaction will fire in the system. When more
coarsely classified reactions are present, however, this is not
always so. In situations where the coarse reactions are
independent of the ES reactions it {\em is\/} safe to increase
$\tau$\/.  Automatically and efficiently differentiating between
this situation and that in which it is not acceptable to increase
$\tau$\/ is not trivial, however, especially when considering
large, complex reaction networks.

Another issue to consider concerns the proper use of
Eq.~(\ref{eq:tau-next}) in our algorithm.  The expression in
(\ref{eq:tau-next}) is a transformation formula in which a new
value of $\tau_\mu^\mathrm{ES}$ is calculated using the new value
of $a_\mu$\/, the old value of $a_\mu$\/, the old value of
$\tau_\mu^\mathrm{ES}$, and the old time step. Under normal
circumstances, the ``old" values are those from the previous
simulation step.  As discussed in note 14 of
Ref.~\onlinecite{Gibson00}, however, this is not always the case.
Specifically, when a reaction becomes inactive, $a_\mu=0$ and
$\tau_\mu^\mathrm{ES} = \infty$. Upon reactivation, these values
clearly cannot be used in Eq.~(\ref{eq:tau-next}). Thus, the
values of $a_\mu$\/, $\tau_\mu^\mathrm{ES}$ and $\tau$\/ are used
from the last simulation step at which $R_\mu$ was active. Values
of $\{a_\nu^\prime(\tau_\nu^{\prime\,\mathrm{ES}} -
\tau^\prime)\}$ are stored for all reactions that did not fire at
the current step and used in Eq.~(\ref{eq:tau-next}) at the next
step at which the corresponding reactions are active. Usually this
is the subsequent step, but sometimes it is not.

In our approach, we must account for the fact that reactions
classified as ES can change their classification in the subsequent
step and then return to an ES classification at a later step.
Although the reaction does not become inactive in this case, the
same procedure of ``carrying over" the values of $a_\mu$\/,
$\tau_\mu^\mathrm{ES}$ and $\tau$\/ is used. We simply extend the
procedure described above, therefore, and use the stored values of
$\{a_\nu^\prime(\tau_\nu^{\prime\,\mathrm{ES}} - \tau^\prime)\}$
at the next step at which the corresponding reactions are active
{\em and\/} are classified as ES.

We also consider the problem of negative populations. We avoid
this problem by simply tracking the state of the system and
reversing the population updates if any populations are found to
be negative after all reaction firings have been accounted for.
The value of $\tau$\/ is then reduced (which is always acceptable)
and all reactions reclassified. We reduce $\tau$\/ by 50\%, but
any amount should suffice.  A reduction in $\tau$\/ will result in
smaller numbers of reaction firings and, hopefully, alleviation of
all negative populations.  If not, then the procedure is repeated
until no further adjustments are necessary.~\cite{note:ES-iter}

This approach is equivalent to the simple ``try again" procedure
discussed and implemented in Ref.~\onlinecite{Cao05:negPop} as a
second layer of protection against the occurrence of negative
populations.  This is used in conjunction with the primary
strategy of identifying critical reactions.  Note, however, that
the critical reaction strategy essentially amounts to a
partitioning of reactions into ES and ``Poisson" classifications
with the intent of avoiding negative populations by maintaining a
fine-level description of reactions with small reactant
populations. Our method already accomplishes this via inclusion of
the ES classification and thus avoids the need to introduce
additional calculations or parameters (see Sec.~\ref{sec:discuss}
for a further discussion).

Finally, in determining the set of reaction firings, use of
Eqs.~(\ref{eq:K-norm}) and (\ref{eq:K-determ}) for ``Langevin" and
``Deterministic" reactions, respectively, will result in values
that are real numbers rather than integers. Since it is difficult
to determine at what point a continuous population description is
acceptable in lieu of an integer description, we choose to round
these values before updating the species populations. In
Ref.~\onlinecite{Gillesp01} it was argued that use of
Eq.~(\ref{eq:K-norm}) as opposed to (\ref{eq:K-poisson}) is an
improvement computationally since generation of normal random
numbers is faster than Poisson random numbers.  Some of this
improvement is clearly negated, therefore, by including a
subsequent rounding operation, although we have yet to
quantitatively investigate the extent of this effect. While the
same argument holds for ``Deterministic" reactions, the
elimination of the random number generation operation should more
than compensate for the added rounding procedure.

With these issues accounted for, the partitioned leaping algorithm
(PLA) is presented as follows:
\begin{enumerate}
    \item[1.]{Initialize (species populations, rate
              constants, define $\approx\!1$, $\gg\!1$,
              $\epsilon$\/, etc.).}
    \item[2.]{Determine the {\em initial\/} value of $\tau$\/
              (see Sec.~\ref{subsec:PLA-timestep}).}
    \item[3.]{Classify all reactions (not already classified as ES)
              using the criteria presented above.}
    \item[4.]{For all (newly classified) ES reactions,
              calculate tentative next-reaction times,
              $\tau_\mu^\mathrm{ES}$, using
              Eqs.~(\ref{eq:tau-first}) and (\ref{eq:tau-next}).}
    \item[5a.]{If $\mathrm{Min}\{\tau_\nu^\mathrm{ES}\} \ne
              \tau$\/ {\em and\/} all reactions are ES, set $\tau
              = \mathrm{Min}\{\tau_\nu^\mathrm{ES}\}$.}
    \item[5b.]{If $\mathrm{Min}\{\tau_\nu^\mathrm{ES}\} < \tau$\/,
              set $\tau = \mathrm{Min}\{\tau_\nu^\mathrm{ES}\}$
              and return to step 3.}
    \item[6.]{Determine the set of reaction firings
              $\{k_\nu(\tau)\}$ using the appropriate
              formulas~\cite{note:deviates}
              and update the species populations.}
    \item[7.]{If any $X_i(t+\tau) < 0$, reverse all population
              updates, set $\tau=\tau/2$ and return to step 3.
              If not, advance the clock by $\tau$\/ and return to
              step 2 if stopping criterion not met.}
\end{enumerate}

It is important to recognize the minimal user intervention
required to implement this approach. Once the reaction network is
defined and the associated rate parameters set, one need only
define three model-{\em independent\/} parameters quantifying the
concepts $\approx\!1$ (for ES classification), $\gg\!1$ (for
coarse classifications) and $\ll\!1$ (defining $\epsilon$\/).  The
algorithm then automatically and dynamically partitions the
reactions into various subsets, correctly accounting for
stochastic noise and ``leaping" over unimportant reaction events.

\subsection{Determining the Initial Time Step} \label{subsec:PLA-timestep}

In practice, there are two alternate approaches for carrying out
Step~2 of the PLA. The first is a reaction-based approach in which
values of $\tau_\mu^\mathrm{Leap}$ are calculated for each
reaction in the system using the constraint expression in
(\ref{eq:constraint}) directly.~\cite{Gillesp01, Gillesp03,
Cao06:newStep}  $\tau$\/ is then set equal to
$\mathrm{Min}\{\tau_\nu^\mathrm{Leap}\}$.  Older versions did this
with $\xi \equiv a_0$ in (\ref{eq:constraint}),~\cite{Gillesp01,
Gillesp03} but the current approach uses $\xi \equiv
a_\mu$.~\cite{Cao06:newStep}   In this approach,
$\tau_\mu^\mathrm{Leap}$ is given by~\cite{Cao06:newStep}
\begin{eqnarray}
    \tau_\mu^\mathrm{Leap} & = & \mathrm{Min}
    \left\{
        \frac{\epsilon_\mu}{|m_\mu(t)|} ,
        \frac{\epsilon_\mu^2}{\sigma_\mu^2(t)}
    \right\}, \label{eq:tauLeap} \\
    \epsilon_\mu & \equiv & \mathrm{Max}\{\epsilon
    a_\mu,\beta_\mu\} \nonumber \\
    m_\mu(t) & \equiv & \sum_{\nu=1}^M f_{\mu\nu}(t) a_\nu(t),
    \nonumber \\
    \sigma_\mu^2(t) & \equiv & \sum_{\nu=1}^M f_{\mu\nu}^2(t)
    a_\nu(t), \nonumber \\
    f_{\mu\nu}(t) & \equiv & \sum_{j=1}^N z_{\nu{j}}
    \frac{\partial a_\mu(t)}{\partial X_j}. \nonumber
\end{eqnarray}

Here, $\beta_\mu$\/ represents the {\em minimum\/} possible change
in the propensity $a_\mu$\/.  This quantity is used to overcome
problems associated with small reactant populations.  In
Ref.~\onlinecite{Cao06:newStep}, $\beta_\mu \equiv c_\mu$\/, the
stochastic rate constant for $R_\mu$\/. For first-order reactions
this definition is exact.  For higher-order reactions, however,
this is a lower bound: If the propensity changes at all it will do
so by an amount $\ge c_\mu$\/.~\cite{Cao06:newStep} Thus, this
approach is simple but may also be restrictive.  We propose an
alternative: An {\em approximate\/} but less restrictive value of
$\beta_\mu$\/ is the smallest {\em non-zero\/} value of
$\{\partial a_\mu /\partial X_j\}$, where $j$\/ indexes all
reactant species involved in $R_\mu$\/.  If all values of
$\{\partial a_\mu /\partial X_j\}$ are zero, however, then
$\beta_\mu = c_\mu$\/.

The second approach for determining $\tau$\/ constrains the
relative changes in the {\em species populations\/} in such a way
that (\ref{eq:constraint}) is satisfied for all
reactions.~\cite{Cao06:newStep}  $\tau$\/ is then set equal to
$\mathrm{Min}\{T_j^\mathrm{Leap}\}$, where~\cite{Cao06:newStep}
\begin{eqnarray}
    T_i^\mathrm{Leap} & = & \mathrm{Min}
    \left\{
        \frac{e_i}{|\hat{m}_i(t)|} ,
        \frac{e_i^2}{\hat{\sigma}_i^2(t)}
    \right\}, \label{eq:teeLeap} \\
    e_i & \equiv & \mathrm{Max}\{\epsilon X_i/g_i, 1\} \nonumber \\
    \hat{m}_i(t) & \equiv & \sum_{\nu=1}^M z_{\nu{i}} a_\nu(t),
    \nonumber \\
    \hat{\sigma}_i^2(t) & \equiv & \sum_{\nu=1}^M z_{\nu{i}}^2
    a_\nu(t). \nonumber
\end{eqnarray}
The parameter $g_i$\/ depends on the highest-order reaction that
species $S_i$\/ is involved in and can be determined by simple
inspection during initialization.~\cite{Cao06:newStep}  The
expressions in (\ref{eq:teeLeap}) require fewer computational
operations than those in (\ref{eq:tauLeap}) and each
$T_i^\mathrm{Leap}$ calculation should thus be significantly
faster than each $\tau_\mu^\mathrm{Leap}$.

In Ref.~\onlinecite{Cao06:newStep}, expressions for $g_i$\/ are
presented for all reaction types up to third order.  For reactions
involving more than one entity of a certain species, such as $2S_i
\rightarrow products$\/, $g_i$\/ is a function of $X_i$\/.  These
expressions have clear upper and lower bounds, however.  Thus, in
order to simplify the calculations we use the more restrictive,
but computationally simpler, upper bounds. We determine $g_i$\/ as
follows:
\begin{center}
\begin{tabular}{@{}r@{\hspace{60pt}}l@{}}
    \multicolumn{1}{l}{If $S_i$\/ is a reactant in \ldots} & then $g_i =$ \ldots \\ \hline
    \hfill $3S_i$ \hfill $\rightarrow products$,        & 11/2, \hfill \textit{else}\/ \ldots \\
    \hfill $2S_i + S_j$ \hfill $\rightarrow products$,  & 9/2,  \hfill \textit{else}\/ \ldots \\
    $ \left\{ \begin{array}{@{}r@{}}
            \hfill S_i + 2S_j \hfill \rightarrow products, \\
            \hfill S_i + S_j + S_k \hfill \rightarrow products, \\
            \hfill 2S_i \hfill \rightarrow products,
            \end{array} \right.\! $                     & 3,    \hfill \textit{else}\/ \ldots \\
    \hfill $S_i + S_j$ \hfill $\rightarrow products$,   & 2,    \hfill \textit{else}\/ \ldots \\
    \hfill $S_i$ \hfill $\rightarrow products$,         & 1,    \hfill \textit{else}\/ \ldots \\
    \multicolumn{1}{@{\hspace{30pt}}l}{zero reactions,} & 0
\end{tabular}
\end{center}
Note that the last item will result in $T_i^\mathrm{Leap} =
\infty$, which will never be $\mathrm{Min}\{T_j^\mathrm{Leap}\}$.

\section{Examples} \label{sec:examples}

With the presentation of our simulation approach now complete, we
will demonstrate in this section the utility of the method, in
terms of efficiency and accuracy, via two illustrative examples.
We will begin by considering a simple model of clustering that
illustrates the algorithm's ability to treat systems in which
species populations vary over many orders of magnitude. A
biologically inspired model system will then be considered that
illustrates how the stochastic process of gene expression can be
accurately and efficiently simulated in conjunction with reactions
involving large reactant populations (e.g., metabolic processes).

\subsection{Simple clustering} \label{subsec:cluster}

Clustering processes are inherently multiscale since large numbers
of small clusters generally coexist within a system with small
numbers of large clusters.  As such, clustering provides an ideal
way to demonstrate the ability of the PLA to treat systems in
which species populations vary over many orders of magnitude.

We have considered a simple clustering model comprised of the
following nine reactions:
\settowidth{\fill}{$R_2:\,\,$ \hfill $S_1 + S_2$}%
\begin{eqnarray}
    \makebox[\fill]{$R_1:\,\,$ \hfill $2S_1$}       & \rightarrow & S_2 \label{ex:cluster} \\
    \makebox[\fill]{$R_2:\,\,$ \hfill $S_1 + S_2$}  & \rightarrow & S_3 \nonumber \\
    \makebox[\fill]{$R_3:\,\,$ \hfill $S_1 + S_3$}  & \rightarrow & S_4 \nonumber \\
    \vdots & & \nonumber \\
    \makebox[\fill]{$R_9:\,\,$ \hfill $S_1 + S_9$}  & \rightarrow & S_{10} \nonumber
\end{eqnarray}
For the sake of simplicity we have neglected dissociation
reactions and assume that monomers ($S_1$) are the only mobile
species in the system. Furthermore, in order to confine the
multiscale effects to variations in the species populations alone,
we have chosen deterministic rate constants such that their
stochastic counterparts are equivalent for all reactions. For
$R_1$ we choose $3.0 \times 10^6$~M$^{-1}$~s$^{-1}$, and for all
other reactions $6.0 \times 10^6$~M$^{-1}$~s$^{-1}$. We set the
initial monomer concentration to $1.66 \times 10^{-6}$~M and
consider various system volumes $\Omega$ ranging from $10^{-15}$
to $10^{-9}$~L. This corresponds to initial monomer populations
$X_1(0) = 10^3$ to $10^9$ and stochastic rate constants $\{c_\nu\}
= 10^{-2}$ to $10^{-8}$~s$^{-1}$.  All simulations were run until
consumption of all monomers was complete.

\begin{figure}
\includegraphics[width=180pt]{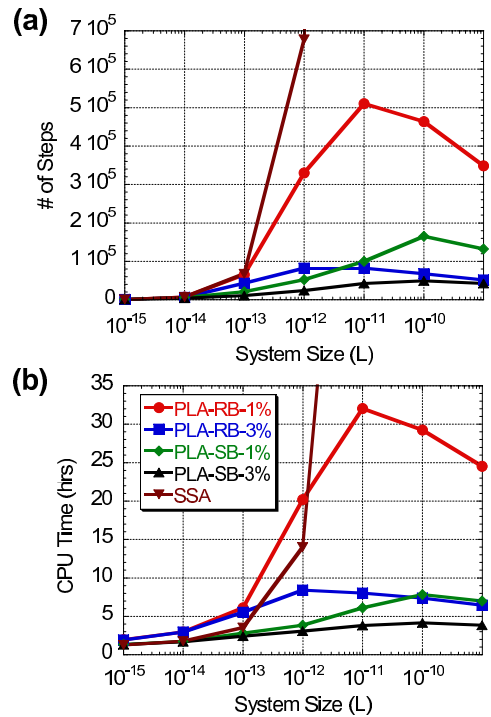}%
\caption{Average numbers of steps (a) and CPU times (b) required
for $10\,000$ PLA and SSA simulation runs of the simple clustering
model (\ref{ex:cluster}) [PLA-RB-1\% means reaction-based
$\tau$\/-selection with $\epsilon=0.01$, etc.]. Reaction
classifications were made in the PLA runs using
$\approx\!{1}\!=\!3$ and $\gg\!{1}\!=\!100$. All simulations were
performed on a 1.80~GHz Athlon processor.
\label{fig:Clust-steps-CPU}}%
\end{figure}

In Fig.~\ref{fig:Clust-steps-CPU}, we compare average numbers of
simulation steps and CPU times required for PLA and SSA
simulations of (\ref{ex:cluster}) at all system sizes considered.
In general, the CPU times follow the same trends as the simulation
steps and the reaction-based calculations (PLA-RB) require more
steps, and more CPU time, than the species-based (PLA-SB).  We
also see that for the smallest system size the SSA and PLA give
identical results, meaning that all reactions were classified as
ES at all steps of the PLA simulations.  As the system size
increases, however, increased amounts of leaping are observed. The
effect is modest for system sizes of $10^{-14}$ and $10^{-13}$~L,
but increases dramatically beyond that. Moreover, the number of
steps actually {\em decreases\/} for PLA simulations of systems
larger than $\sim\!10^{-11}$~L.  The effects of leaping thus
overtake the system size effects for these large systems.

\begin{figure}[t]
\includegraphics[width=240pt]{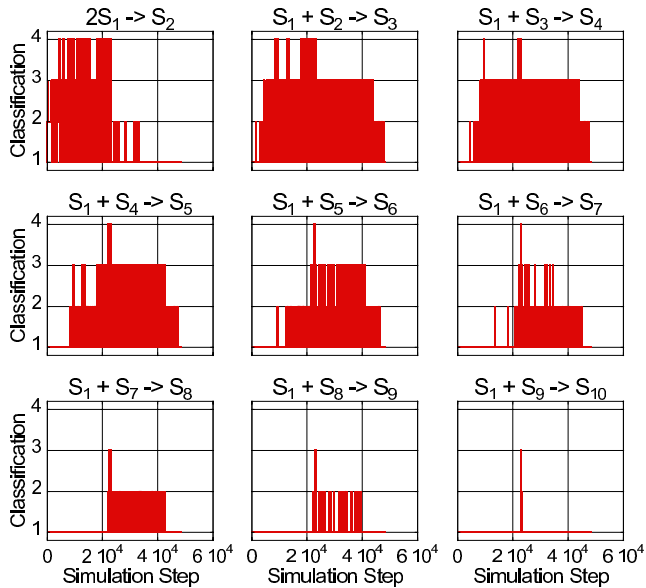}%
\caption{Classifications vs.\ simulation step for each reaction of
the simple clustering model (\ref{ex:cluster}) at $\Omega =
10^{-9}$~L [i.e., $X_1(0) = 10^9$].  Classifications are: (1)
Exact Stochastic, (2) Poisson, (3) Langevin, (4) Deterministic.
Results are for a single PLA-RB-3\% simulation using
$\approx\!{1}\!=\!3$ and $\gg\!{1}\!=\!100$.
\label{fig:Clust-classif-1e9}}%
\end{figure}

In Fig.~\ref{fig:Clust-classif-1e9}, we show the classifications
achieved for each reaction in (\ref{ex:cluster}) at each
simulation step of a single PLA-RB-3\% simulation at a system size
of $10^{-9}$~L. These results show leaping in action and
illustrate the inherent multiscale nature of the reaction network.
Reactions involving the smallest clusters (i.e., $R_1$--$R_4$)
experience extensive amounts of leaping for much of the
simulation, with classifications varying rapidly between ES,
``Poisson," ``Langevin" and, at times, ``Deterministic."
Conversely, reactions involving larger clusters experience much
less leaping. Reactions $R_7$--$R_9$, for example, experience only
small amounts of leaping up into the ``Langevin" regime.

As for accuracy, smoothed frequency histograms (see Appendix) were
generated from $10\,000$ simulation runs of the SSA and the PLA
for various cluster sizes (the populations of which vary by as
many as four orders of magnitude) at various system sizes.
Modified versions of the histogram distance $D$\/ and the self
distance $D^\mathrm{self}_\mathrm{SSA}$, introduced by Cao and
Petzold,~\cite{Cao06:Stats} were then used to make quantitative
comparisons. In all cases, the calculated histogram distances were
smaller than the SSA self distances, indicating excellent accuracy
(see the Appendix for further details).

\subsection{Stochastic gene expression} \label{subsec:geneExpr}

The origins and consequences of stochasticity in biological
systems has been a subject of intense interest
recently.~\cite{McAdams97, Arkin98, McAdams99, Elowitz02, Fedo02,
Rao02, Kaern05, Raser05} In cellular systems, the primary source
of ``intrinsic" stochastic noise is gene expression, where the
small numbers of regulatory molecules involved in the process
result in proteins being produced in ``bursts" rather than
continuously.~\cite{McAdams97, McAdams99, Kaern05} Other cellular
processes, such as metabolism, often involve large numbers of
molecules and it has been shown that stochastic fluctuations in
gene expression can quantitatively affect these
dynamics.~\cite{Puch04} A fully stochastic treatment of biological
systems involving both gene expression and metabolism is
infeasible,~\cite{Endy01} however, and thus provides motivation
for developing multiscale simulation methods capable of treating
systems involving both large- and small-number dynamics.

We have applied the PLA to a simple biologically inspired model
system that involves both gene expression and protein-protein
interactions.  The network that we have considered is as follows:
\settowidth{\fill}{$R_3:\,\,$ \hfill $\{P\!:\!Q\}$}%
\begin{eqnarray}
    \makebox[\fill]{$R_1:\,\,$ \hfill $G$}           & \rightarrow & G^\ast      \label{ex:geneExpr} \\
    \makebox[\fill]{$R_2:\,\,$ \hfill $G^\ast$}      & \rightarrow & G + nP      \nonumber \\
    \makebox[\fill]{$R_3:\,\,$ \hfill $P + Q$}       & \rightarrow & \{P\!:\!Q\} \nonumber \\
    \makebox[\fill]{$R_4:\,\,$ \hfill $\{P\!:\!Q\}$} & \rightarrow & R + Q       \nonumber \\
    \makebox[\fill]{$R_5:\,\,$ \hfill $R$}           & \rightarrow & \ast        \nonumber
\end{eqnarray}
The first two reactions constitute the gene expression part of the
network, where the single gene $G$\/ spontaneously converts into
an active conformation $G^\ast$\/ that then produces proteins
$P$\/ in bursts of $n$\/. The third and fourth reactions
constitute the protein-protein enzymatic part of the network where
$P$\/ interacts with $Q$\/ to form an enzyme-substrate complex
$\{P\!:\!Q\}$ that then produces $R$\/ and regenerates $Q$\/. The
final reaction models the degradation of $R$\/.

Rate constants for the five reactions in (\ref{ex:geneExpr}) were
set equal to 750~s$^{-1}$, 750~s$^{-1}$,
$6.0\!\times\!10^8$~M$^{-1}$~s$^{-1}$, 100~s$^{-1}$, and
50~s$^{-1}$, respectively.  The initial enzyme concentration
$[Q]_\mathrm{o}$ was set equal to $1.66\!\times\!10^{-7}$~M, $n$\/
to $0.2X_Q(0)$,~\cite{note:burst} and investigations were carried
out for various system sizes ranging from $10^{-15}$ to
$10^{-7}$~L. This corresponds to initial enzyme populations
$X_Q(0)$ ranging from $10^2$ to $10^{10}$. In all cases, the
system began with a single entity of $G$\/ and null populations of
$G^\ast$\/, $\{P\!:\!Q\}$ and $R$\/. All simulations were run
until $t=1$~s.

\begin{figure}[t]
\includegraphics[width=240pt]{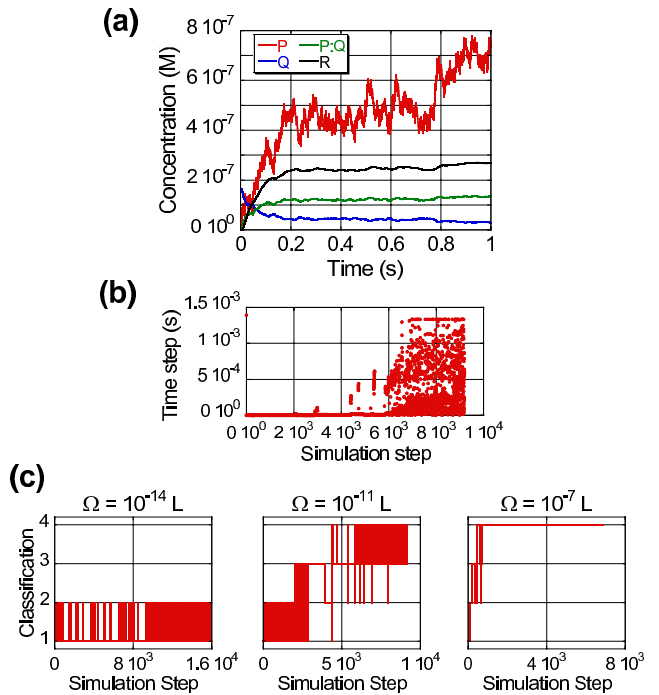}%
\caption{Results for typical PLA-RB-3\% simulations
($\approx\!{1}\!=\!3$, $\gg\!{1}\!=\!100$) of the simple gene
expression model (\ref{ex:geneExpr}): (a) A typical time course at
$\Omega = 10^{-11}$~L, (b) value of the time step at each
simulation step for the same run as in (a), (c) classifications
vs.\ simulation step for reaction $R_3$ at various system sizes.
\label{fig:GeneExpr-conc_tau-1e6}}
\end{figure}

In Fig.~\ref{fig:GeneExpr-conc_tau-1e6}, we show typical results
for PLA-RB-3\% simulations of (\ref{ex:geneExpr}). The time course
plot in Fig.~\ref{fig:GeneExpr-conc_tau-1e6}(a), generated for a
system size of $10^{-11}$~L, illustrates the stochastic nature of
the gene expression dynamics, apparent in the noisy time evolution
of the gene product $P$\/. Conversely, the large-number dynamics
result in much smoother trajectories for $Q$\/, $\{P\!:\!Q\}$ and
$R$\/. The time step plot in
Fig.~\ref{fig:GeneExpr-conc_tau-1e6}(b), taken from the same
simulation as in \ref{fig:GeneExpr-conc_tau-1e6}(a), illustrates
the algorithm's ability to dynamically adjust as the system
evolves. The first 3000 simulation steps correspond to
$\sim\!0.01$~s of simulated time, while a time of $0.1$~s is
reached around step 6400. The entire simulation takes $9192$ steps
to complete. Thus, simulating the first 1\% of the evolved time
required $\sim\!33\%$ of all simulation steps and the first 10\%
$\sim\!70\%$ of all steps.

In Fig.~\ref{fig:GeneExpr-conc_tau-1e6}(c), we see how the
classifications coarsen with increasing system size.  For
$\Omega\!=\!10^{-14}$~L the classifications for $R_3$ never reach
beyond ``Poisson."  For $\Omega\!=\!10^{-11}$~L the
classifications coarsen approximately midway through the
simulation, and for $\Omega\!=\!10^{-7}$~L ``Deterministic" status
is achieved quickly and maintained almost exclusively throughout.
Similar results are obtained for reactions $R_4$ and $R_5$ (data
not shown) while $R_1$ and $R_2$ are classified as ES at all steps
of the simulation.

\begin{figure}
\includegraphics[width=180pt]{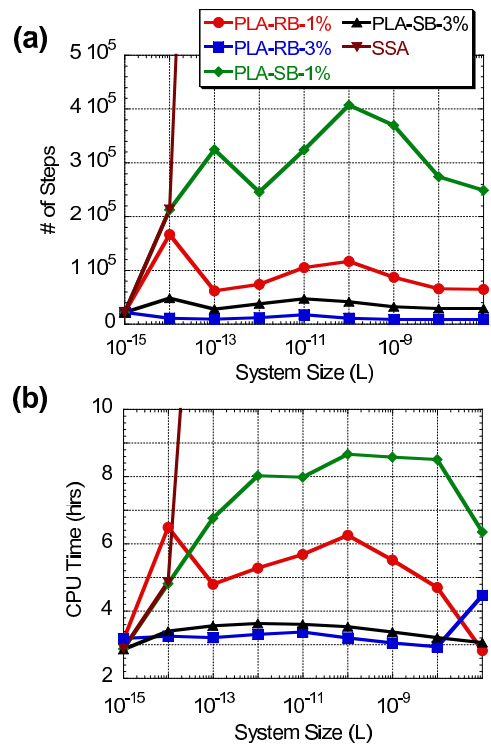}%
\caption{Average numbers of steps (a) and CPU times (b) required
for $10\,000$ PLA and SSA simulation runs of the simple gene
expression model (\ref{ex:geneExpr}). Reaction classifications
were made in the PLA runs using $\approx\!{1}\!=\!3$ and
$\gg\!{1}\!=\!100$. All simulations were performed on a 1.80~GHz
Athlon processor.
\label{fig:GeneExpr-steps-CPU}}%
\end{figure}

In Fig.~\ref{fig:GeneExpr-steps-CPU}, we compare the average
numbers of simulation steps and the CPU times required for PLA and
SSA simulations of (\ref{ex:geneExpr}).  Again we see that the CPU
times generally follow the same trends as the simulation steps
but, contrary to what is observed in Sec.~\ref{subsec:cluster},
the SB calculations require more steps, and longer CPU times, than
the RB.  The efficiencies of the two $\tau$\/-selection procedures
thus appear to be system-dependent. This is an interesting result
and an area of possible future investigation.

We also see in Fig.~\ref{fig:GeneExpr-steps-CPU} that the general
trend for the simulation steps is an initial increase with system
size, followed by a drop, another increase and then another drop
with an eventual leveling off (the only departure from this is for
the PLA-RB-3\% where the number of steps initially decreases with
system size).  The initial behavior is the same as that seen for
the simple clustering model (see Fig.~\ref{fig:Clust-steps-CPU})
and can easily be explained in terms of a competition between
system size and leaping effects. The second increase is unique to
this example, however, and arises from an increasing fraction of
simulations requiring large numbers of steps.  At $10^{-10}$~L,
for example, 95\% of all PLA-RB-1\% simulations require between
38\,078 and 437\,061 steps to complete. At $10^{-12}$~L, the range
is 38\,081 and 142\,687. The stochastic nature of the gene
expression dynamics thus results in a wide variety of possible
time-evolution trajectories for systems of size $10^{-11}$ to
$10^{-9}$~L, some of which require many more simulation steps to
complete than others. The ability of the PLA to handle systems in
this ``medium" size range is a particular strength of the method.

Finally, in terms of accuracy, we generated smoothed frequency
histograms for each species in (\ref{ex:geneExpr}) at various
system sizes and again observed excellent agreement between the
PLA and SSA. In all cases, the calculated histogram distances were
smaller than the SSA self distances.

\section{Discussion and Conclusions} \label{sec:discuss}

In this article, we have presented a new multiscale simulation
approach, termed the ``partitioned leaping algorithm," for
modeling stochasticity in chemical reaction networks. Clearly,
this method is closely related to
$\tau$\/-leaping.~\cite{Gillesp01, Gillesp03, Cao05:negPop,
Cao06:newStep} The key difference is that reactions are classified
individually in the PLA, which leads to various advantages over
$\tau$\/-leaping: The PLA overcomes the problem of negative
populations more simply, the SSA is incorporated into the
algorithmic framework in a more seamless and natural way and
reactions are treated at four levels of description rather than
two.

The negative populations problem is avoided in the PLA primarily
through the structure of the algorithm.  This can be seen most
easily by considering the RB $\tau$\/-selection procedure
(\ref{eq:tauLeap}), though the arguments hold for the SB approach
as well. If a reaction has a small reactant population, such that
the reaction would be deemed ``critical" in
$\tau$\/-leaping,~\cite{Cao05:negPop, Cao06:newStep} then its
$\tau_\mu^\mathrm{Leap}$ value will be $\leq\!1/a_\mu$\/, the
expected time of its next firing.  This is because, for very small
reactant populations, a single firing will change $a_\mu$\/ by a
fractional amount $\geq\!\epsilon$\/, which will be automatically
recognized in the $\tau$\/-selection process. As a result, the
reaction will be classified as ES (since $\tau$\/ is always
$\leq\!\tau_\mu^\mathrm{Leap}$) and thus prevented from firing
more than once, assuring that its reactant populations do not fall
below zero. Since this outcome arises as a natural consequence of
the design of the PLA, there is no need for additional
calculations or parameters, such as those associated with the
critical reaction search in $\tau$\/-leaping.~\cite{Cao05:negPop,
Cao06:newStep}

Another difference between the methods is that in $\tau$\/-leaping
the $\tau$\/-selection formulas (\ref{eq:tauLeap}) and
(\ref{eq:teeLeap}) are applied only to non-critical
reactions.~\cite{Cao05:negPop, Cao06:newStep} In the PLA, these
formulas are applied to all reactions. This is merely a
superficial difference, however. Although critical reactions are
treated separately from non-critical reactions in
$\tau$\/-leaping, they still contribute to $\tau$\/-selection,
just in a different way.  In $\tau$\/-leaping, ``candidate" time
steps are calculated for the critical and non-critical reaction
subsets and $\tau$\/ set equal to the smaller of the
two.~\cite{Cao05:negPop, Cao06:newStep}  The critical reaction
time step is calculated using Eq.~(\ref{eq:Direct-tau}) with $a_0$
replaced by $a_0^c$\/, the sum of the critical reaction
propensities.  Obviously, this will give a time interval on the
order of $1/\mathrm{Max}\{a_\nu\}$. In the PLA, these same
reactions will be treated using Eqs.~(\ref{eq:tauLeap}) or
(\ref{eq:teeLeap}). As just discussed, in the PLA-RB scenario the
result will be $\tau_\mu^\mathrm{Leap}$ values on the order of
$1/a_\mu$\/. $\tau$\/ is set equal to
$\mathrm{Min}\{\tau_\nu^\mathrm{Leap}\}$, so the overall result is
essentially the same: the ``critical" reactions contribute to the
$\tau$\/-selection process a time interval on the order of
$1/\mathrm{Max}\{a_\nu\}$.  The randomness associated with these
reactions is then added in in Step~4 of the PLA using the methods
of the NRM~\cite{Gibson00} [Eqs.~(\ref{eq:tau-first}) and
(\ref{eq:tau-next})], which is the obvious choice given the
structure of the algorithm.

The ability of the PLA to treat reactions at the
continuous-stochastic and deterministic scales, rather than just
the discrete-stochastic, is an important attribute of the method
as well; the PLA could be seen as unifying, into one overarching
approach, three different types of simulation methods:
exact-stochastic,~\cite{Gillesp76, Gibson00}
$\tau$\/-leaping,~\cite{Gillesp01, Cao06:newStep} and methodology
coupling, or hybrid, techniques.~\cite{Hasel02, Burr04, Puch04,
Vasu04, Kiehl04, Salis05:hybrid}  Hybrid methods operate by
partitioning reactions into ``fast" and ``slow" subsets and using,
e.g., deterministic reaction rate equations or stochastic
differential equations, to describe the fast reactions and the SSA
(or modified versions thereof) for the slow.  The primary
shortcoming of these approaches is the lack of a sound theoretical
basis for the partitioning scheme,~\cite{Cao04:stability} which
calls into question the extent of their utility.  The partitioning
procedure described in Sec.~\ref{subsec:PLA-algo} overcomes this
problem; partitioning is accomplished via Gillespie's rigorously
derived criteria for transitioning between the descriptions
(\ref{eq:K-poisson})--(\ref{eq:K-determ}).~\cite{Gillesp00,
Gillesp01} Of course, as currently formulated, ``Langevin" and
``Deterministic" reactions are treated in a manner equivalent to
the explicit Euler method for solving stochastic and ordinary
differential equations, respectively, which may not be
satisfactory in all cases.

Recognizing this, Petzold and co-workers have developed various
``implicit" $\tau$\/-leaping methods~\cite{Rath03,
Cao04:stability, Rath05} that take into account the values of the
propensities at both the beginning {\em and\/} end of the time
step $\tau$\/.  By doing so, these methods are able to take larger
time steps than explicit methods~\cite{Gillesp01, Gillesp03} that
only consider the propensity at the beginning of the step (the
algorithm presented here is explicit). This comes at a cost,
however: Implicit methods dampen fluctuations in the species
populations and, hence, underestimate the amplitude of the
noise.~\cite{Rath03, Cao04:stability, Rath05}  Nevertheless, the
ability of these methods to maintain stability in situations where
explicit methods fail has been demonstrated.  Incorporating these
ideas into the PLA is an interesting possibility, and an area of
possible future investigation.

From a {\em practical\/} point of view, it is unclear to what
extent the efficiency of the PLA is actually improved by including
the ``Langevin" and ``Deterministic" classifications. As mentioned
above, the reason for including these is to presumably improve
computational efficiency via faster generation of normal random
numbers~\cite{Gillesp01} or by eliminating random number
generation altogether.  Their inclusion adds complexity to the
method, however. To what extent this attenuates the gains in
efficiency remains to be seen, and is something we plan on
investigating further in the future. Note, however, that a useful
feature of the PLA is its ability, via manipulation of the
classification parameters, to force a simulation at a particular
level of description (e.g., if $\gg\!1,
\approx\!1\!\equiv\!\infty$, then all reactions will be classified
as ES at all steps of a simulation -- See
Sec.~\ref{subsec:PLA-algo}). With the coarse classifications
included, one could use this feature (with rounding turned off) to
perform, e.g., deterministic simulations of a system for direct
comparisons to results from the PLA.

Finally, we must also acknowledge the primary shortcoming of the
PLA, namely, in handling systems with widely disparate rate
constants.  Consider a system that contains a single entity, such
as a gene, experiencing rapid ON/OFF or binding/unbinding
behavior. Because there exists only one entity these reactions
will be flagged as ES, and because of the large rate constants
their $\tau_\mu^\mathrm{Leap}$ values (in the PLA-RB) will be
small.  The result will be small time steps and a ``bogging down"
of the PLA. Note that this is a problem in $\tau$\/-leaping as
well.

The only way to overcome this problem is to ``reduce" the model so
that the effects of the fast reactions are accounted for in some
approximate way.  This is a common approach; examples include
quasi-steady-state or Michaelis-Menten reductions. Recently, more
advanced model reduction schemes have been proposed that can be
applied to generalized reaction networks~\cite{Rao03,
Cao05:slowSSA, Cao05:MSSA, Gout05} and, in some cases,
dynamically.~\cite{Samant05, Salis05:QSSA} Since these methods are
specifically designed to overcome problems associated with widely
disparate rate constants they could form an effective complement
to leaping techniques that tackle the problem of widely disparate
species populations. Future techniques combining automatic and
dynamic model reduction with partitioned leaping could open the
door to computational investigations far beyond the reach of
current approaches.

\begin{acknowledgments}
F.~A.\ Escobedo, H. Lee, A.~A.\ Quong, A.~J.\ Golumbfskie and
C.~F.\ Melius are thanked for useful discussions regarding this
work. We acknowledge financial support from the Semiconductor
Research Corporation Graduate Fellowship Program.
\end{acknowledgments}

\appendix*
\section{Histogram smoothing, histogram distance and
self distance}

For a set of $N$\/ data points $\{x_1, x_2, \ldots, x_N\}$, the
total number falling within a discrete interval $[x, x+\Delta)$
can be formally written as $\sum_{i=1}^N \int_x^{x+\Delta}
\delta(x_i-x^\prime) dx^\prime$, where $\delta(x_i-x^\prime)$ is
the Dirac delta function and the integral equals unity if $x_i$\/
lies within $[x, x+\Delta)$ and zero otherwise.  A ``histogram
density" can then be obtained by dividing this quantity by
$N\Delta$ and taking the limit as $\Delta \rightarrow 0$,
\begin{equation}
    \hat{h}(x) = \lim_{\Delta \rightarrow 0} \frac{1}{N\Delta}
                 \sum_{i=1}^N \int_x^{x+\Delta}
                 \delta(x_i-x^\prime) dx^\prime.%
                 \label{eq:h-hat}
\end{equation}
The ``hat" in $\hat{h}(x)$ signifies that this quantity is a
statistical {\em estimator\/} of the true histogram density
\begin{equation}
    h(x) = \lim_{N \rightarrow \infty} \hat{h}(x).%
    \label{eq:h-true}
\end{equation}

\begin{figure}
\includegraphics[width=240pt]{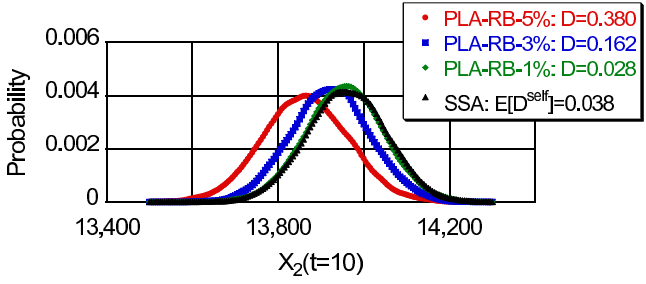}%
\caption{Smoothed population histograms for species $S_2$ at
$t\!=\!10$ obtained from $10\,000$ simulation runs of the
``decaying-dimerizing" reaction set~\cite{Gillesp03} using the SSA
and the PLA-RB with various values of $\epsilon$\/. Reaction
classifications were made in the PLA runs using
$\approx\!{1}\!=\!3$ and $\gg\!{1}\!=\!100$.  All histograms were
generated using Eq.~(\ref{eq:h-hat-approx}) with $\sigma=15$.
\label{fig:DD-X2}}%
\end{figure}

A ``smoothed" histogram can then be obtained by approximating the
delta function in (\ref{eq:h-hat}) by a finite width Gaussian
$\kappa \exp\left(\frac{-(x_i-x^\prime)^2}{2\sigma^2}\right)$,
where $\kappa$ is a normalization constant and $\sigma^2$ is the
(user-defined) variance.  Substituting into (\ref{eq:h-hat}),
noting that to first order $\int_x^{x+\Delta} e^{-u^2} du = \Delta
e^{-x^2}$, and normalizing, gives
\begin{equation}
    \hat{h}(x) \cong \frac{1}{\sqrt{2\pi} \sigma N} \sum_{i=1}^N
    \exp\left(\frac{-(x_i-x)^2}{2\sigma^2}\right).%
    \label{eq:h-hat-approx}
\end{equation}
All smoothed histograms generated in this work were obtained using
this expression.  Examples are shown in Fig.~\ref{fig:DD-X2},
obtained from 10\,000 PLA and SSA simulations of the
``decaying-dimerizing" reaction set using the same rate parameters
and initial conditions as in Ref.~\onlinecite{Gillesp03}.

In order to quantitatively compare results obtained via different
simulation methods (i.e., PLA, SSA, etc.)\ we use the ``histogram
distance" discussed by Cao and Petzold.~\cite{Cao06:Stats}  The
quantity $\delta{h_x}$\/ is defined as $h_1(x)-h_2(x)$ and the
histogram distance $D$\/ is then
\begin{equation}
    D \equiv \frac{1}{2} \int_x |\delta{h_x}| dx.%
    \label{eq:D}
\end{equation}
The factor of 1/2 assures that $D$\/ lies within [0,1), with 0
constituting a perfect fit and 1 a complete mismatch.

It is important to note that $D$\/ is defined in (\ref{eq:D}) in
terms of the {\em true\/} histogram densities $h_1(x)$ and
$h_2(x)$.  In practice, we only have their estimators and can thus
only calculate an estimated value of $D$.  As a result, a certain
amount of statistical uncertainty is associated with the
comparison of histograms.  In order to quantify this uncertainty
Cao and Petzold~\cite{Cao06:Stats} introduce the ``self distance,"
$D^\mathrm{self}$, which essentially measures the distance between
the estimator $\hat{h}(x)$ and the true density $h(x)$.
Expressions for the upper bounds on the mean and variance of
$D^\mathrm{self}$ are presented in Ref.~\onlinecite{Cao06:Stats}
in terms of the number of bin intervals $K$\/.  However, since we
are using Eq.~(\ref{eq:h-hat-approx}) rather than a counting
procedure to generate $\hat{h}(x)$ we must derive alternate
expressions.

We begin by defining
\begin{eqnarray}
    D^\mathrm{self} & \equiv & \frac{1}{2} \int_x
    |\delta{h_x^\mathrm{self}}| dx, \\
    \delta{h_x^\mathrm{self}} & \equiv & \hat{h}(x) - h(x).
    \nonumber%
    \label{eq:Dself}
\end{eqnarray}
Following Cao and Petzold,~\cite{Cao06:Stats} we then note that
the number of points falling within the interval $[x, x+\Delta)$
is a binomial random variable $B(p_x,N)$, where $p_x$\/ is the
success probability. $\hat{h}(x)$ can then be written as
\begin{equation}
    \hat{h}(x) = \lim_{\Delta \rightarrow 0} B(p_x,N)/N\Delta = B(dp_x,N)/Ndx.%
    \label{eq:h-hat-binom}
\end{equation}
Since the mean E$[B(p_x,N)] = Np_x$\/ and the variance
Var$[B(p_x,N)] = Np_xq_x$ ($q_x \equiv 1-p_x$), the mean and
variance of $\hat{h}(x)$ are $dp_x/dx$ and $dp_xdq_x/Ndx^2$,
respectively. With $dp_x=h(x)dx$\/ and $dq_x=1-h(x)dx$\/, the mean
and variance of $\delta h_x^\mathrm{self}$ are then
\begin{eqnarray}
    \mathrm{E}[\delta h_x^\mathrm{self}] & = & 0, \\
    \mathrm{var}[\delta h_x^\mathrm{self}] & = &
    [h(x)dx][1-h(x)dx]/Ndx^2 \approx \hat{h}(x)/Ndx, \nonumber%
\end{eqnarray}
where the last line utilizes the histogram density estimator
$\hat{h}(x)$.

For large $N$\/, $\delta h_x^\mathrm{self}$ can be approximated as
a normal random variable with mean zero and variance
$\hat{h}(x)/Ndx$\/.  This means that
\begin{equation}
    \frac{\delta h_x^\mathrm{self}}
            {\sqrt{\hat{h}(x)/Ndx}}
    \label{eq:stand-norm}
\end{equation}
is approximately standard normal and
\begin{equation}
    \frac{|\delta h_x^\mathrm{self}|}
         {\sqrt{\hat{h}(x)/Ndx}}
    \label{eq:chi}
\end{equation}
is approximately chi distributed with one degree of freedom. Since
the mean and variance of a chi random variable with one degree of
freedom are $\sqrt{2/\pi}$ and $(\pi-2)/\pi$, respectively, this
gives
\begin{eqnarray}
    \mathrm{E}[|\delta h_x^\mathrm{self}|] & \approx &
    \sqrt{\frac{2}{N\pi} \frac{\hat{h}(x)}{dx}}, \\
    \mathrm{var}[|\delta h_x^\mathrm{self}|] & \approx &
    \left(\frac{\pi-2}{N\pi}\right) \frac{\hat{h}(x)}{dx}.
    \nonumber%
    \label{eq:abs-delta-hx}
\end{eqnarray}
Finally, using the identities~\cite{Gillesp-book}
\begin{eqnarray}
    \textstyle%
    \mathrm{E}\left[\int_x f(x)dx\right] & = & \int_x
    \mathrm{E}[f(x)]dx, \nonumber \\
    \textstyle%
    \mathrm{Var}\left[\int_x f(x)dx\right] & \leq & \left( \int_x
    \sqrt{\mathrm{Var}[f(x)]}dx\right)^2, \nonumber
\end{eqnarray}
we get
\begin{eqnarray}
    \mathrm{E}[D^\mathrm{self}] & \approx & \frac{1}{2}
    \sqrt{\frac{2}{N\pi}} \int_x \sqrt{\hat{h}(x)dx},  \\
    \mathrm{var}[D^\mathrm{self}] & \lesssim & \frac{1}{4}
    \left(\frac{\pi-2}{N\pi}\right) \left( \int_x \sqrt{
    \hat{h}(x)dx} \right)^2. \nonumber
\end{eqnarray}

In practice, we calculate
$\mathrm{E}[D^\mathrm{self}_\mathrm{ref}]$, the self distance for
a {\em reference\/} histogram generally obtained using the SSA.
This value then tells us that any histogram with a distance $D <
\mathrm{E}[D^\mathrm{self}_\mathrm{ref}]$ {\em cannot\/} be
statistically differentiated from the reference histogram.  In
Fig.~\ref{fig:DD-X2}, we see that only the PLA-RB-1\% histogram
achieves this level of accuracy. The expression for
$\mathrm{var}[D^\mathrm{self}]$ is included here for completeness.

\newpage


\begin{thebibliography}{48}

\expandafter\ifx\csname
natexlab\endcsname\relax\def\natexlab#1{#1}\fi
\expandafter\ifx\csname bibnamefont\endcsname\relax
  \def\bibnamefont#1{#1}\fi
\expandafter\ifx\csname bibfnamefont\endcsname\relax
  \def\bibfnamefont#1{#1}\fi
\expandafter\ifx\csname citenamefont\endcsname\relax
  \def\citenamefont#1{#1}\fi
\expandafter\ifx\csname url\endcsname\relax
  \def\url#1{\texttt{#1}}\fi
\expandafter\ifx\csname
urlprefix\endcsname\relax\def\urlprefix{URL }\fi
\providecommand{\bibinfo}[2]{#2}
\providecommand{\eprint}[2][]{\url{#2}}

\bibitem[{\citenamefont{McAdams and Arkin}(1997)}]{McAdams97}
\bibinfo{author}{\bibfnamefont{H.~H.} \bibnamefont{McAdams}} \bibnamefont{and}
  \bibinfo{author}{\bibfnamefont{A.}~\bibnamefont{Arkin}},
  \bibinfo{journal}{Proc. Natl. Acad. Sci. USA} \textbf{\bibinfo{volume}{94}},
  \bibinfo{pages}{814} (\bibinfo{year}{1997}).

\bibitem[{\citenamefont{Arkin et~al.}(1998)\citenamefont{Arkin, Ross, and
  McAdams}}]{Arkin98}
\bibinfo{author}{\bibfnamefont{A.~P.} \bibnamefont{Arkin}},
  \bibinfo{author}{\bibfnamefont{J.}~\bibnamefont{Ross}}, \bibnamefont{and}
  \bibinfo{author}{\bibfnamefont{H.~H.} \bibnamefont{McAdams}},
  \bibinfo{journal}{Genetics} \textbf{\bibinfo{volume}{149}},
  \bibinfo{pages}{1633} (\bibinfo{year}{1998}).

\bibitem[{\citenamefont{McAdams and Arkin}(1999)}]{McAdams99}
\bibinfo{author}{\bibfnamefont{H.~H.} \bibnamefont{McAdams}} \bibnamefont{and}
  \bibinfo{author}{\bibfnamefont{A.}~\bibnamefont{Arkin}},
  \bibinfo{journal}{Trends Genet.} \textbf{\bibinfo{volume}{15}},
  \bibinfo{pages}{65} (\bibinfo{year}{1999}).

\bibitem[{\citenamefont{Elowitz et~al.}(2002)\citenamefont{Elowitz, Levine,
  Siggia, and Swain}}]{Elowitz02}
\bibinfo{author}{\bibfnamefont{M.~B.} \bibnamefont{Elowitz}},
  \bibinfo{author}{\bibfnamefont{A.~J.} \bibnamefont{Levine}},
  \bibinfo{author}{\bibfnamefont{E.~D.} \bibnamefont{Siggia}},
  \bibnamefont{and} \bibinfo{author}{\bibfnamefont{P.~S.} \bibnamefont{Swain}},
  \bibinfo{journal}{Science} \textbf{\bibinfo{volume}{297}},
  \bibinfo{pages}{1183} (\bibinfo{year}{2002}).

\bibitem[{\citenamefont{Fedoroff and Fontana}(2002)}]{Fedo02}
\bibinfo{author}{\bibfnamefont{N.}~\bibnamefont{Fedoroff}} \bibnamefont{and}
  \bibinfo{author}{\bibfnamefont{W.}~\bibnamefont{Fontana}},
  \bibinfo{journal}{Science} \textbf{\bibinfo{volume}{297}},
  \bibinfo{pages}{1129} (\bibinfo{year}{2002}).

\bibitem[{\citenamefont{Rao et~al.}(2002)\citenamefont{Rao, Wolf, and
  Arkin}}]{Rao02}
\bibinfo{author}{\bibfnamefont{C.~V.} \bibnamefont{Rao}},
  \bibinfo{author}{\bibfnamefont{D.~M.} \bibnamefont{Wolf}}, \bibnamefont{and}
  \bibinfo{author}{\bibfnamefont{A.~P.} \bibnamefont{Arkin}},
  \bibinfo{journal}{Nature} \textbf{\bibinfo{volume}{420}},
  \bibinfo{pages}{231} (\bibinfo{year}{2002}).

\bibitem[{\citenamefont{K{\ae}rn et~al.}(2005)\citenamefont{K{\ae}rn, Elston,
  Blake, and Collins}}]{Kaern05}
\bibinfo{author}{\bibfnamefont{M.}~\bibnamefont{K{\ae}rn}},
  \bibinfo{author}{\bibfnamefont{T.~C.} \bibnamefont{Elston}},
  \bibinfo{author}{\bibfnamefont{W.~J.} \bibnamefont{Blake}}, \bibnamefont{and}
  \bibinfo{author}{\bibfnamefont{J.~J.} \bibnamefont{Collins}},
  \bibinfo{journal}{Nature Rev. Genet.} \textbf{\bibinfo{volume}{6}},
  \bibinfo{pages}{451} (\bibinfo{year}{2005}).

\bibitem[{\citenamefont{Raser and O'Shea}(2005)}]{Raser05}
\bibinfo{author}{\bibfnamefont{J.~M.} \bibnamefont{Raser}} \bibnamefont{and}
  \bibinfo{author}{\bibfnamefont{E.~K.} \bibnamefont{O'Shea}},
  \bibinfo{journal}{Science} \textbf{\bibinfo{volume}{309}},
  \bibinfo{pages}{2010} (\bibinfo{year}{2005}).

\bibitem[{\citenamefont{Plummer and Griffin}(1995)}]{Plumm95}
\bibinfo{author}{\bibfnamefont{J.~D.} \bibnamefont{Plummer}} \bibnamefont{and}
  \bibinfo{author}{\bibfnamefont{P.~B.} \bibnamefont{Griffin}},
  \bibinfo{journal}{Nucl. Instr. Meth. Phys. Res. B}
  \textbf{\bibinfo{volume}{102}}, \bibinfo{pages}{160} (\bibinfo{year}{1995}).

\bibitem[{\citenamefont{Roy and Asenov}(2005)}]{Roy05}
\bibinfo{author}{\bibfnamefont{S.}~\bibnamefont{Roy}} \bibnamefont{and}
  \bibinfo{author}{\bibfnamefont{A.}~\bibnamefont{Asenov}},
  \bibinfo{journal}{Science} \textbf{\bibinfo{volume}{309}},
  \bibinfo{pages}{388} (\bibinfo{year}{2005}).

\bibitem[{ITR()}]{ITRS01}
\bibinfo{note}{The International Technology Roadmap for Semiconductors, 2001
  Ed., http://public.itrs.net}.

\bibitem[{\citenamefont{Gillespie}(1976)}]{Gillesp76}
\bibinfo{author}{\bibfnamefont{D.~T.} \bibnamefont{Gillespie}},
  \bibinfo{journal}{J. Comput. Phys.} \textbf{\bibinfo{volume}{22}},
  \bibinfo{pages}{403} (\bibinfo{year}{1976}).

\bibitem[{\citenamefont{Gibson and Bruck}(2000)}]{Gibson00}
\bibinfo{author}{\bibfnamefont{M.~A.} \bibnamefont{Gibson}} \bibnamefont{and}
  \bibinfo{author}{\bibfnamefont{J.}~\bibnamefont{Bruck}}, \bibinfo{journal}{J.
  Phys. Chem. A} \textbf{\bibinfo{volume}{104}}, \bibinfo{pages}{1876}
  (\bibinfo{year}{2000}).

\bibitem[{\citenamefont{Cao et~al.}(2004{\natexlab{a}})\citenamefont{Cao, Li,
  and Petzold}}]{Cao04:efficientSSA}
\bibinfo{author}{\bibfnamefont{Y.}~\bibnamefont{Cao}},
  \bibinfo{author}{\bibfnamefont{H.}~\bibnamefont{Li}}, \bibnamefont{and}
  \bibinfo{author}{\bibfnamefont{L.}~\bibnamefont{Petzold}},
  \bibinfo{journal}{J. Chem. Phys.} \textbf{\bibinfo{volume}{121}},
  \bibinfo{pages}{4059} (\bibinfo{year}{2004}{\natexlab{a}}).

\bibitem[{\citenamefont{Resat et~al.}(2001)\citenamefont{Resat, Wiley, and
  Dixon}}]{Resat01}
\bibinfo{author}{\bibfnamefont{H.}~\bibnamefont{Resat}},
  \bibinfo{author}{\bibfnamefont{H.~S.} \bibnamefont{Wiley}}, \bibnamefont{and}
  \bibinfo{author}{\bibfnamefont{D.~A.} \bibnamefont{Dixon}},
  \bibinfo{journal}{J. Phys. Chem. B} \textbf{\bibinfo{volume}{105}},
  \bibinfo{pages}{11026} (\bibinfo{year}{2001}).

\bibitem[{\citenamefont{Gillespie}(2001)}]{Gillesp01}
\bibinfo{author}{\bibfnamefont{D.~T.} \bibnamefont{Gillespie}},
  \bibinfo{journal}{J. Chem. Phys.} \textbf{\bibinfo{volume}{115}},
  \bibinfo{pages}{1716} (\bibinfo{year}{2001}).

\bibitem[{\citenamefont{Gillespie and Petzold}(2003)}]{Gillesp03}
\bibinfo{author}{\bibfnamefont{D.~T.} \bibnamefont{Gillespie}}
  \bibnamefont{and} \bibinfo{author}{\bibfnamefont{L.~R.}
  \bibnamefont{Petzold}}, \bibinfo{journal}{J. Chem. Phys.}
  \textbf{\bibinfo{volume}{119}}, \bibinfo{pages}{8229} (\bibinfo{year}{2003}).

\bibitem[{\citenamefont{Rathinam et~al.}(2003)\citenamefont{Rathinam, Petzold,
  Cao, and Gillespie}}]{Rath03}
\bibinfo{author}{\bibfnamefont{M.}~\bibnamefont{Rathinam}},
  \bibinfo{author}{\bibfnamefont{L.~R.} \bibnamefont{Petzold}},
  \bibinfo{author}{\bibfnamefont{Y.}~\bibnamefont{Cao}}, \bibnamefont{and}
  \bibinfo{author}{\bibfnamefont{D.~T.} \bibnamefont{Gillespie}},
  \bibinfo{journal}{J. Chem. Phys.} \textbf{\bibinfo{volume}{119}},
  \bibinfo{pages}{12784} (\bibinfo{year}{2003}).

\bibitem[{\citenamefont{Cao et~al.}(2004{\natexlab{b}})\citenamefont{Cao,
  Petzold, Rathinam, and Gillespie}}]{Cao04:stability}
\bibinfo{author}{\bibfnamefont{Y.}~\bibnamefont{Cao}},
  \bibinfo{author}{\bibfnamefont{L.~R.} \bibnamefont{Petzold}},
  \bibinfo{author}{\bibfnamefont{M.}~\bibnamefont{Rathinam}}, \bibnamefont{and}
  \bibinfo{author}{\bibfnamefont{D.~T.} \bibnamefont{Gillespie}},
  \bibinfo{journal}{J. Chem. Phys.} \textbf{\bibinfo{volume}{121}},
  \bibinfo{pages}{12169} (\bibinfo{year}{2004}{\natexlab{b}}).

\bibitem[{\citenamefont{Cao et~al.}(2005{\natexlab{a}})\citenamefont{Cao,
  Gillespie, and Petzold}}]{Cao05:negPop}
\bibinfo{author}{\bibfnamefont{Y.}~\bibnamefont{Cao}},
  \bibinfo{author}{\bibfnamefont{D.~T.} \bibnamefont{Gillespie}},
  \bibnamefont{and} \bibinfo{author}{\bibfnamefont{L.~R.}
  \bibnamefont{Petzold}}, \bibinfo{journal}{J. Chem. Phys.}
  \textbf{\bibinfo{volume}{123}}, \bibinfo{pages}{054104}
  (\bibinfo{year}{2005}{\natexlab{a}}).

\bibitem[{\citenamefont{Rathinam et~al.}(2005)\citenamefont{Rathinam, Petzold,
  Cao, and Gillespie}}]{Rath05}
\bibinfo{author}{\bibfnamefont{M.}~\bibnamefont{Rathinam}},
  \bibinfo{author}{\bibfnamefont{L.~R.} \bibnamefont{Petzold}},
  \bibinfo{author}{\bibfnamefont{Y.}~\bibnamefont{Cao}}, \bibnamefont{and}
  \bibinfo{author}{\bibfnamefont{D.~T.} \bibnamefont{Gillespie}},
  \bibinfo{journal}{Multiscale Model. Simul.} \textbf{\bibinfo{volume}{4}},
  \bibinfo{pages}{867} (\bibinfo{year}{2005}).

\bibitem[{\citenamefont{Cao et~al.}(2006)\citenamefont{Cao, Gillespie, and
  Petzold}}]{Cao06:newStep}
\bibinfo{author}{\bibfnamefont{Y.}~\bibnamefont{Cao}},
  \bibinfo{author}{\bibfnamefont{D.~T.} \bibnamefont{Gillespie}},
  \bibnamefont{and} \bibinfo{author}{\bibfnamefont{L.~R.}
  \bibnamefont{Petzold}}, \bibinfo{journal}{J. Chem. Phys.}
  \textbf{\bibinfo{volume}{124}}, \bibinfo{pages}{044109}
  (\bibinfo{year}{2006}).

\bibitem[{\citenamefont{Tian and Burrage}(2004)}]{Tian04}
\bibinfo{author}{\bibfnamefont{T.}~\bibnamefont{Tian}} \bibnamefont{and}
  \bibinfo{author}{\bibfnamefont{K.}~\bibnamefont{Burrage}},
  \bibinfo{journal}{J. Chem. Phys.} \textbf{\bibinfo{volume}{121}},
  \bibinfo{pages}{10356} (\bibinfo{year}{2004}).

\bibitem[{\citenamefont{Chatterjee et~al.}(2005)\citenamefont{Chatterjee,
  Vlachos, and Katsoulakis}}]{Chatt05}
\bibinfo{author}{\bibfnamefont{A.}~\bibnamefont{Chatterjee}},
  \bibinfo{author}{\bibfnamefont{D.~G.} \bibnamefont{Vlachos}},
  \bibnamefont{and} \bibinfo{author}{\bibfnamefont{M.~A.}
  \bibnamefont{Katsoulakis}}, \bibinfo{journal}{J. Chem. Phys.}
  \textbf{\bibinfo{volume}{122}}, \bibinfo{pages}{024112}
  (\bibinfo{year}{2005}).

\bibitem[{\citenamefont{Haseltine and Rawlings}(2002)}]{Hasel02}
\bibinfo{author}{\bibfnamefont{E.~L.} \bibnamefont{Haseltine}}
  \bibnamefont{and} \bibinfo{author}{\bibfnamefont{J.~B.}
  \bibnamefont{Rawlings}}, \bibinfo{journal}{J. Chem. Phys.}
  \textbf{\bibinfo{volume}{117}}, \bibinfo{pages}{6959} (\bibinfo{year}{2002}).

\bibitem[{\citenamefont{Burrage et~al.}(2004)\citenamefont{Burrage, Tian, and
  Burrage}}]{Burr04}
\bibinfo{author}{\bibfnamefont{K.}~\bibnamefont{Burrage}},
  \bibinfo{author}{\bibfnamefont{T.}~\bibnamefont{Tian}}, \bibnamefont{and}
  \bibinfo{author}{\bibfnamefont{P.}~\bibnamefont{Burrage}},
  \bibinfo{journal}{Prog. Biophys. Mol. Biol.} \textbf{\bibinfo{volume}{85}},
  \bibinfo{pages}{217} (\bibinfo{year}{2004}).

\bibitem[{\citenamefont{Pucha{\l}ka and Kierzek}(2004)}]{Puch04}
\bibinfo{author}{\bibfnamefont{J.}~\bibnamefont{Pucha{\l}ka}} \bibnamefont{and}
  \bibinfo{author}{\bibfnamefont{A.~M.} \bibnamefont{Kierzek}},
  \bibinfo{journal}{Biophys. J.} \textbf{\bibinfo{volume}{86}},
  \bibinfo{pages}{1357} (\bibinfo{year}{2004}).

\bibitem[{\citenamefont{Vasudeva and Bhalla}(2004)}]{Vasu04}
\bibinfo{author}{\bibfnamefont{K.}~\bibnamefont{Vasudeva}} \bibnamefont{and}
  \bibinfo{author}{\bibfnamefont{U.~S.} \bibnamefont{Bhalla}},
  \bibinfo{journal}{Bioinformatics} \textbf{\bibinfo{volume}{20}},
  \bibinfo{pages}{78} (\bibinfo{year}{2004}).

\bibitem[{\citenamefont{Kiehl et~al.}(2004)\citenamefont{Kiehl, Mattheyses, and
  Simmons}}]{Kiehl04}
\bibinfo{author}{\bibfnamefont{T.~R.} \bibnamefont{Kiehl}},
  \bibinfo{author}{\bibfnamefont{R.~M.} \bibnamefont{Mattheyses}},
  \bibnamefont{and} \bibinfo{author}{\bibfnamefont{M.~K.}
  \bibnamefont{Simmons}}, \bibinfo{journal}{Bioinformatics}
  \textbf{\bibinfo{volume}{20}}, \bibinfo{pages}{316} (\bibinfo{year}{2004}).

\bibitem[{\citenamefont{Salis and
  Kaznessis}(2005{\natexlab{a}})}]{Salis05:hybrid}
\bibinfo{author}{\bibfnamefont{H.}~\bibnamefont{Salis}} \bibnamefont{and}
  \bibinfo{author}{\bibfnamefont{Y.}~\bibnamefont{Kaznessis}},
  \bibinfo{journal}{J. Chem. Phys.} \textbf{\bibinfo{volume}{122}},
  \bibinfo{pages}{054103} (\bibinfo{year}{2005}{\natexlab{a}}).

\bibitem[{\citenamefont{Rao and Arkin}(2003)}]{Rao03}
\bibinfo{author}{\bibfnamefont{C.~V.} \bibnamefont{Rao}} \bibnamefont{and}
  \bibinfo{author}{\bibfnamefont{A.~P.} \bibnamefont{Arkin}},
  \bibinfo{journal}{J. Chem. Phys.} \textbf{\bibinfo{volume}{118}},
  \bibinfo{pages}{4999} (\bibinfo{year}{2003}).

\bibitem[{\citenamefont{Cao et~al.}(2005{\natexlab{b}})\citenamefont{Cao,
  Gillespie, and Petzold}}]{Cao05:slowSSA}
\bibinfo{author}{\bibfnamefont{Y.}~\bibnamefont{Cao}},
  \bibinfo{author}{\bibfnamefont{D.~T.} \bibnamefont{Gillespie}},
  \bibnamefont{and} \bibinfo{author}{\bibfnamefont{L.~R.}
  \bibnamefont{Petzold}}, \bibinfo{journal}{J. Chem. Phys.}
  \textbf{\bibinfo{volume}{122}}, \bibinfo{pages}{014116}
  (\bibinfo{year}{2005}{\natexlab{b}}).

\bibitem[{\citenamefont{Cao et~al.}(2005{\natexlab{c}})\citenamefont{Cao,
  Gillespie, and Petzold}}]{Cao05:MSSA}
\bibinfo{author}{\bibfnamefont{Y.}~\bibnamefont{Cao}},
  \bibinfo{author}{\bibfnamefont{D.}~\bibnamefont{Gillespie}},
  \bibnamefont{and} \bibinfo{author}{\bibfnamefont{L.}~\bibnamefont{Petzold}},
  \bibinfo{journal}{J. Comput. Phys.} \textbf{\bibinfo{volume}{206}},
  \bibinfo{pages}{395} (\bibinfo{year}{2005}{\natexlab{c}}).

\bibitem[{\citenamefont{Goutsias}(2005)}]{Gout05}
\bibinfo{author}{\bibfnamefont{J.}~\bibnamefont{Goutsias}},
  \bibinfo{journal}{J. Chem. Phys.} \textbf{\bibinfo{volume}{122}},
  \bibinfo{pages}{184102} (\bibinfo{year}{2005}).

\bibitem[{\citenamefont{Samant and Vlachos}(2005)}]{Samant05}
\bibinfo{author}{\bibfnamefont{A.}~\bibnamefont{Samant}} \bibnamefont{and}
  \bibinfo{author}{\bibfnamefont{D.~G.} \bibnamefont{Vlachos}},
  \bibinfo{journal}{J. Chem. Phys.} \textbf{\bibinfo{volume}{123}},
  \bibinfo{pages}{144114} (\bibinfo{year}{2005}).

\bibitem[{\citenamefont{Salis and
  Kaznessis}(2005{\natexlab{b}})}]{Salis05:QSSA}
\bibinfo{author}{\bibfnamefont{H.}~\bibnamefont{Salis}} \bibnamefont{and}
  \bibinfo{author}{\bibfnamefont{Y.~N.} \bibnamefont{Kaznessis}},
  \bibinfo{journal}{J. Chem. Phys.} \textbf{\bibinfo{volume}{123}},
  \bibinfo{pages}{214106} (\bibinfo{year}{2005}{\natexlab{b}}).

\bibitem[{\citenamefont{Gillespie}(2000)}]{Gillesp00}
\bibinfo{author}{\bibfnamefont{D.~T.} \bibnamefont{Gillespie}},
  \bibinfo{journal}{J. Chem. Phys.} \textbf{\bibinfo{volume}{113}},
  \bibinfo{pages}{297} (\bibinfo{year}{2000}).

\bibitem[{not({\natexlab{a}})}]{note:subscr}
\bibinfo{note}{The notation used in this article differs from that in
  Refs.~\onlinecite{Gillesp01, Gillesp03, Rath03, Cao04:stability,
  Cao05:negPop, Rath05, Cao06:newStep} and \onlinecite{Gillesp00}. Here, latin
  character subscripts are used to refer to species and greek characters to
  reactions}.

\bibitem[{\citenamefont{Gillespie}(1992{\natexlab{a}})}]{Gillesp92}
\bibinfo{author}{\bibfnamefont{D.~T.} \bibnamefont{Gillespie}},
  \bibinfo{journal}{Physica A} \textbf{\bibinfo{volume}{188}},
  \bibinfo{pages}{404} (\bibinfo{year}{1992}{\natexlab{a}}).

\bibitem[{not({\natexlab{b}})}]{note:label}
\bibinfo{note}{The superscript ``ES" is used because differentiating between
  exact-stochastic time intervals and ``leaping" intervals will be important
  later}.

\bibitem[{not({\natexlab{c}})}]{note:NRM-differ}
\bibinfo{note}{The expression in Eq.~\protect(\ref{eq:tau-next}) differs
  slightly from that in Ref.~\onlinecite{Gibson00} as it is a {\em relative\/}
  time version of the transformation formula}.

\bibitem[{not({\natexlab{d}})}]{note:ES-iter}
\bibinfo{note}{The number of iterations required in this procedure is
  definitively finite. In extreme situations, if $\tau$\/ is continually
  reduced, at some point {\em all\/} reactions will become classified as ES.
  Reclassifications will then no longer be necessary and the iterative loop
  will terminate}.

\bibitem[{not({\natexlab{e}})}]{note:deviates}
\bibinfo{note}{Standard techniques exist for generating Poisson and normal
  random deviates.~\cite{NumRec} For ES reactions, if $\tau_\mu^\mathrm{ES} =
  \tau$\/ then $k_\mu(\tau) = 1$, otherwise zero}.

\bibitem[{\citenamefont{Cao and Petzold}(2006)}]{Cao06:Stats}
\bibinfo{author}{\bibfnamefont{Y.}~\bibnamefont{Cao}} \bibnamefont{and}
  \bibinfo{author}{\bibfnamefont{L.}~\bibnamefont{Petzold}},
  \bibinfo{journal}{J. Comput. Phys.} \textbf{\bibinfo{volume}{212}},
  \bibinfo{pages}{6} (\bibinfo{year}{2006}).

\bibitem[{\citenamefont{Endy and Brent}(2001)}]{Endy01}
\bibinfo{author}{\bibfnamefont{D.}~\bibnamefont{Endy}} \bibnamefont{and}
  \bibinfo{author}{\bibfnamefont{R.}~\bibnamefont{Brent}},
  \bibinfo{journal}{Nature} \textbf{\bibinfo{volume}{409}},
  \bibinfo{pages}{391} (\bibinfo{year}{2001}).

\bibitem[{not({\natexlab{f}})}]{note:burst}
\bibinfo{note}{By allowing the number of proteins produced per expression event
  to change we are effectively varying the degree of ``translational
  efficiency"~\cite{Kaern05}}.

\bibitem[{\citenamefont{Gillespie}(1992{\natexlab{b}})}]{Gillesp-book}
\bibinfo{author}{\bibfnamefont{D.~T.} \bibnamefont{Gillespie}},
  \emph{\bibinfo{title}{Markov Processes: An Introduction for Physical
  Scientists}} (\bibinfo{publisher}{Academic, San Diego},
  \bibinfo{year}{1992}{\natexlab{b}}).

\bibitem[{\citenamefont{\mbox{See, e.g.} et~al.}(1999)\citenamefont{\mbox{See,
  e.g.}, Press, Teukolsky, Vetterling, and Flannery}}]{NumRec}
\bibinfo{author}{\bibnamefont{\mbox{See, e.g.}}},
  \bibinfo{author}{\bibfnamefont{W.~H.} \bibnamefont{Press}},
  \bibinfo{author}{\bibfnamefont{S.~A.} \bibnamefont{Teukolsky}},
  \bibinfo{author}{\bibfnamefont{W.~T.} \bibnamefont{Vetterling}},
  \bibnamefont{and} \bibinfo{author}{\bibfnamefont{B.~P.}
  \bibnamefont{Flannery}}, \emph{\bibinfo{title}{Numerical Recipes in C, The
  Art of Scientific Computing, 2nd Ed.}} (\bibinfo{publisher}{Cambridge
  University Press, New York, NY}, \bibinfo{year}{1999}).

\end{thebibliography}

\end{document}